\documentclass[iop]{emulateapj}
\usepackage{epstopdf}

\slugcomment{draft version January 13, 2017, accepted for publication in the Astrophysical Journal}

\shorttitle{3C\,368 CO-Dark Star Formation}
\shortauthors{Lamarche et al.}

\begin{document}

\title{CO-Dark Star Formation and Black Hole Activity in 3C\,368 at $z = 1.131$: \\
Coeval Growth of Stellar and Supermassive Black Hole Masses}

\author{C. Lamarche\altaffilmark{1}, G. Stacey\altaffilmark{1}, D. Brisbin\altaffilmark{2}, C. Ferkinhoff\altaffilmark{3}, S. Hailey-Dunsheath\altaffilmark{4}, T. Nikola\altaffilmark{5}, D. Riechers\altaffilmark{1}, C. E. Sharon\altaffilmark{6},  H. Spoon\altaffilmark{5}, and A. Vishwas\altaffilmark{1}}

\email{cjl272@cornell.edu}
\altaffiltext{1}{Department of Astronomy, Cornell University,
    Ithaca, NY 14853.}
\altaffiltext{2}{N\'{u}cleo de Astronom\'{\i}a, Facultad de Ingenier\'{\i}a y Ciencias, Universidad Diego Portales, Av. Ej\'{e}rcito 441, Santiago, Chile.}
\altaffiltext{3}{Department of Physics, Winona State University, Winona, MN, 55987.}
\altaffiltext{4}{California Institute of Technology, Mail Code 301-17, 1200 East California Boulevard, Pasadena, CA 91125.}
\altaffiltext{5}{Cornell Center for Astrophysics and Planetary Science, Cornell University, Ithaca, NY 14853.}
\altaffiltext{6}{Department of Physics \& Astronomy, McMaster University, 1280 Main Street West, Hamilton, ON L85-4M1, Canada.}

\begin{abstract}
We present the detection of four far-infrared fine-structure oxygen lines, as well as strong upper limits for the CO(2--1) and [N\,{\sc ii}] 205\,$\micron$ lines, in 3C\,368, a well-studied radio-loud galaxy at $z = 1.131$. These new oxygen lines, taken in conjunction with previously observed neon and carbon fine-structure lines, suggest a powerful active galactic nucleus (AGN), accompanied by vigorous and extended star formation. A starburst dominated by O8 stars, with an age of $\sim$ 6.5 Myr, provides a good fit to the fine-structure line data. This estimated age of the starburst makes it nearly concurrent with the latest episode of AGN activity, suggesting a link between the growth of the supermassive black hole and stellar population in this source. We do not detect the CO(2--1) line, down to a level twelve times lower than the expected value for star forming galaxies. This lack of CO line emission is consistent with recent star formation activity if the star-forming molecular gas has low metallicity, is highly fractionated (such that CO is photodissociated through much of the clouds), or is chemically very young (such that CO has not yet had time to form). It is also possible, though we argue unlikely, that the ensemble of fine structure lines are emitted from the region heated by the AGN.
\end{abstract}


\thanks{\emph{Herschel} is an ESA space observatory with science instruments provided by European-led Principal Investigator consortia and with important participation from NASA.}

\thanks{Based on observations carried out with the IRAM Plateau de Bure Interferometer. IRAM is supported by INSU/CNRS (France), MPG (Germany) and IGN (Spain).}

\section{Introduction}

A major goal in the study of galaxy evolution is understanding the interplay between active galactic nuclei (AGN) and star formation in their host galaxies. A correlation exists between the mass of central black holes, which power AGN, and stellar bulge mass, which suggests that the growth of black holes by accretion is linked to the growth of stellar mass by star formation \citep[e.g.,][]{Ferrarese2000}. Gas accretion onto a central black hole can spur AGN activity, including winds and high energy radiation, which may disrupt the interstellar medium (ISM) of the host galaxy, quenching star formation \citep[e.g.,][]{Sanders1988}. On the other hand, AGN activity, through jets and winds for example, can contribute an additional pressure to the ISM of the host galaxy, leading to cloud collapse, and thereby bolstering star formation \citep[both theoretically and observationally, e.g.][]{Silk2013, Gaibler2012, Croft2006, Dey1997}. Here we investigate the relationship between star formation and AGN activity in the bright radio galaxy 3C\,368 at redshift $z$ $=$ 1.131 \citep[as determined by the {[O\,{\sc ii}]} 372.8\,nm line,][]{Meisenheimer1992}.  We observe 3C\,368 as it was in the epoch of both peak star formation and AGN activity in the Universe.

3C\,368 was discovered as part of the Third Cambridge Radio Catalogue \citep{Edge1959}. Since then, it has become one of the best studied, high-redshift, radio-loud, Fanaroff-Riley class II (FR-II) type galaxies, with observations spanning the electromagnetic spectrum from the radio \citep[e.g.,][]{Best1998} to the X-Ray \citep[e.g.,][]{Crawford1995}.

Not only does it exhibit considerable AGN activity, with radio emission extended over 73\,kpc \citep{Best1998}, 3C\,368 also has a substantial stellar mass \citep[$\sim$ 3.6\,$\times$\,10$^{11}$\,$M_{\odot}$,][]{Best1998stellarmass}, and is undergoing a period of vigorous \citep[350\,$M_\odot$\,yr$^{-1}$, calculated by spectral energy distribution (SED) modeling in the far-infrared,][]{Podigachoski2015} and extended \citep[over $\sim$ 2$-$4\,kpc,][]{Stacey2010CII} star formation. \cite{McCarthy1987} suggest a link between AGN activity and star formation in this source, in that the jets emanating from the AGN seem to be spurring star formation. To date, however, no CO emission, which traces the molecular gas necessary for star formation, has been detected in this source \citep[e.g.,][]{Evans1996}.

To further explore the properties of the stellar population in 3C\,368, including its age and spatial distribution, we observed the [O\,{\sc i}] 63\,$\micron$ line, the [O\,{\sc iii}] 52 and 88\,$\micron$ lines, and the [O\,{\sc iv}] 26\,$\micron$ line using the \emph{PACS} spectrometer on the \emph{Herschel Space Observatory}, the [N\,{\sc ii}] 205\,$\micron$ line with the Atacama Large Millimeter/submillimeter Array (ALMA), and the CO(2--1) line using both ALMA and the Plateau de Bure Interferometer (PdBI). The [O\,{\sc i}] line arises from dense photodissociated gas on the surfaces of molecular clouds, and combined with our prior detection of the [C\,{\sc ii}] 158\,$\micron$ line with ZEUS on the Caltech Submillimeter Observatory (CSO) \citep{Stacey2010CII}, traces the strength of the UV radiation field. The [O\,{\sc iii}] 52 and 88\,$\micron$ lines taken together allow us to estimate the density within H {\sc ii} regions, and the [O\,{\sc iv}] 26\,$\micron$ line, with an ionization potential of 54.93\,eV, traces the narrow line region (NLR) of the AGN.

Taken together with spectroscopic observations conducted with the \emph{Infrared Spectrograph (IRS)} on board the \emph{Spitzer Space Telescope}, including [Ne\,{\sc ii}] and [Ne\,{\sc iii}] lines in the mid-infrared, these new data allow us to estimate the age of the starburst in 3C\,368, furthering the theory of AGN-driven star formation in this powerful, yet enigmatic, galaxy. 

We assume a flat $\Lambda$CDM cosmology, with $\Omega_M$ = 0.27, $\Omega_\Lambda$ = 0.73, and H$_0$ = 71\,km\,s$^{-1}$\,Mpc$^{-1}$, throughout this paper \citep{Spergel2003}.

\section{Observations}

\subsection{Herschel/PACS}

The oxygen fine-structure lines were observed in 3C\,368 using the \emph{Photodetector Array Camera and Spectrometer} (\emph{PACS}) \citep{Poglitsch2010} on board the \emph{Herschel Space Observatory} \citep{Pilbratt2010}. The [O\,{\sc i}] 63\,$\micron$ and [O\,{\sc iii}] 88\,$\micron$ lines were observed in line scan chop-nod mode (Obs.~ID: 1342243546), with a total observing time of $\sim$124\,minutes, before the [O\,{\sc iii}] 52\,$\micron$ and [O\,{\sc iv}] 26\,$\micron$ lines were observed simultaneously in range scan mode (Obs.~ID: 1342243547), with an observing time of $\sim$38\,minutes. All observations took place on March 25, 2012.

These observations were reduced using the Herschel Interactive Processing Environment (HIPE, Version 13) \citep{Ott2010}. For each of the oxygen lines, with the exception of the [O\,{\sc i}] 63\,$\micron$ line, the central 9$\farcs$4\,$\times$\,9$\farcs$4 spatial pixel (spaxel) was used to generate the spectrum, and the results were point source corrected. For the [O\,{\sc i}] 63\,$\micron$ line, the inner 3$\times$3 spaxels of the integral field spectrometer were combined, and point source corrected, to create one spectrum. This step was taken in order to correct what appears to be pointing jitter with this observation, as indicated by the presence of line flux on multiple spaxels. Significant line flux was not seen to extend over multiple spaxels for any of the other lines observed by \emph{PACS}.

These line fluxes are reported in Table 1, and the line spectra are shown in Figure 1.

\subsection{PdBI}

As a follow-up to our detection of the [C\,{\sc ii}] line, we attempted to detect the CO(2--1) rotational line with the Plateau de Bure Interferometer (PdBI) in 3C\,368. The observations were carried out on May 31 and June 1, 2010, with five antennas in the D configuration, producing a synthesized beam of size 4$\farcs$69\,$\times$\,4$\farcs$26 (FWHM). Total on source integration was 5.25\,hours, with the Band-1 receivers tuned to the redshifted frequency of the line, 108.183\,GHz. We used the WideX correlator, which provides an instantaneous 3.6 GHz coverage in both polarizations. For these observations, we used MWC349 for absolute flux calibration. The data were calibrated and imaged using GILDAS\footnote{https://www.iram.fr/IRAMFR/GILDAS}, resulting in a 1$\sigma$ sensitivity of 279\,$\mu$Jy\,beam$^{-1}$ for these observations, calculated over an assumed line width of 500\,km\,s$^{-1}$; to match the width of the [C\,{\sc ii}] line. The line was not detected, with a 3$\sigma$ upper limit, assuming the source is small with respect to the (39\,$\times$\,35\,kpc) beam, of 0.42\,Jy\,km\,s$^{-1}$, or equivalently 1.5\,$\times$\,10$^{-21}$\,W\,m$^{-2}$.

We detect the 1.3\,mm rest-frame continuum in both the Northern and Southern lobes of 3C\,368. We report these values using our subsequent, more sensitive, ALMA observations.

\subsection{ALMA}

Having made no detection with the PdBI, we next attempted to detect the CO(2--1) line in 3C\,368 with the Atacama Large Millimeter/submillimeter Array (ALMA) \footnote{The National Radio Astronomy Observatory is a facility of the National Science Foundation operated under cooperative agreement by Associated Universities, Inc.}. These band 3 observations were performed on January 31, 2015, with a synthesized beam size of 3$\farcs$12\,$\times$\,1$\farcs$81 (FWHM). Our on-source integration time was $\sim$37\,minutes. 

For these observations, which were conducted with a precipitable water vapor (PWV) measurement of 7.9\,$\pm$\,0.1\,mm, we used Titan as a flux calibrator, at a distance of 41.7$^\circ$ from the source, and J1751+0939 as both a phase and bandpass calibrator, located 3.6$^\circ$ from the source. Characteristic absolute flux calibrations with ALMA are accurate to $\sim$ 10\% (ALMA Technical Handbook). The data were reduced, imaged, and cleaned with the Common Astronomy Software Application (CASA)\footnote{https://casa.nrao.edu/}, version 4.3.1. 

The 1$\sigma$ sensitivity for these observations is 134\,$\mu$Jy\,beam$^{-1}$, over an assumed line width of 500\,km\,s$^{-1}$. Again the line is not detected. A 3$\sigma$ detection limit for a source enclosed by the 25.8\,$\times$\,15.0\,kpc beam yields an upper limit on the line intensity of 0.201\,Jy\,km\,s$^{-1}$, or equivalently 7.3\,$\times$\,10$^{-22}$\,W\,m$^{-2}$.

This time, we detect the continuum in three components of 3C\,368: the Northern lobe at 673\,$\pm$\,39\,$\mu$Jy, the Southern lobe at 712\,$\pm$\,45\,$\mu$Jy, and the core at 123\,$\pm$\,34\,$\mu$Jy (see Figure 2).

In addition to the CO(2--1) line, we attempted to detect the [N\,{\sc ii}] 205\,$\micron$ line with ALMA. These band 9 observations were carried out on February 21 and April 13, 2014, with a synthesized beam of size 0$\farcs$41\,$\times$\,0$\farcs$29 (FWHM), and an on-source integration time of $\sim$54\,minutes.

For these observations, which were conducted with a PWV of 0.32\,$\pm$\,0.02\,mm, we used Titan as a flux calibrator, at a distance of 47.9$^\circ$ from the source, J1751+0939 as a phase calibrator, located 3.6$^\circ$ from the source, and J1924-2914 as a bandpass calibrator, 44.77$^\circ$ from the source. Characteristic absolute flux calibrations for observations performed in cycle 1 with the ALMA Band 9 receivers are accurate to $\sim$ 15\% (ALMA Technical Handbook). The data were reduced, imaged, and cleaned with CASA, version 4.2.1.

 We obtain a 1$\sigma$ sensitivity of 891\,$\mu$Jy\,beam$^{-1}$, over an assumed line width of 500\,km\,s$^{-1}$. Presuming that the line source is extended over $\sim$ 2 beams (as the continuum contours in Figure 2 suggest), the 3$\sigma$ upper limit is 3.78\,Jy\,km\,s$^{-1}$ or 8.65\,$\times$\,$10^{-20}$\,W\,m$^{-2}$.

We detect the rest-frame 205\,$\micron$ continuum, which traces thermal dust emission, from only the core component of 3C\,368, with an integrated intensity of 5.6\,$\pm$\,1.3\,mJy (see Figure 2).

\subsection{Spitzer/IRS}

The [Ne\,{\sc ii}] 12.81\,$\micron$, [Ne\,{\sc iii}] 15.56\,$\micron$, and [Ne\,{\sc v}] 14.32\,$\micron$ lines were obtained from the Cornell Atlas of \emph{Spitzer/IRS} Sources (CASSIS, Version 6) \citep{Lebouteiller2011}, and were observed as part of a program to study multiple radio galaxies and quasars near the peak of cosmic star formation using the \emph{Spitzer Space Observatory} (PI: Martin Haas, AORKEY=22912000). The observations of 3C\,368 were carried out on May 1, 2008, with the \emph{InfraRed Spectrograph (IRS)} \citep{Houck2004} in low-resolution, long-slit mode, (LL1), covering a wavelength range of 19.5$-$38\,$\micron$. The total on-source integration time for these observations was $\sim$48\,minutes. The data were reduced and extracted using the CASSIS software, which employs an ``optimal" extraction technique in order to produce a spectrum with the highest possible signal to noise ratio \citep{Lebouteiller2015}.

The line fluxes from the three neon lines are reported in Table 1, and the spectrum is plotted in Figure 3.

\section{Results and Discussion}

\subsection{Line Fluxes}

In order to determine the line fluxes for the fine-structure lines observed with \emph{Herschel/PACS}, we sum the flux from the line channels (shaded in Figure 1), and propagate the statistical uncertainties per channel to calculate the signal to noise ratios. The channels associated with the spectral lines are determined by a Gaussian fit to the line profiles (see Table 1), with two exceptions.

First, the central line velocity and full-width-half-maximum (FWHM) obtained by a Gaussian fit to the [O\,{\sc i}] 63\,$\micron$ line, our most significant \emph{Herschel/PACS} detection, are used to determine the line channels for the [O\,{\sc iii}] 88\,$\micron$ line. This is necessary due to the low signal to noise in the [O\,{\sc iii}] 88\,$\micron$ line spectrum.

Second, a Gaussian fit to the [O\,{\sc iii}] 52\,$\micron$ line shows that it is considerably broader ($\Delta$v $\sim$ 760\,km\,s$^{-1}$) than any of the other lines which we observed in 3C\,368 (green dashed line, Figure 1). The reported flux for this line (see Table 1) is calculated by summing the flux from all channels under this best-fit Gaussian (shaded both yellow and cyan in Figure 1). Given that the signal to noise per channel in the [O\,{\sc iii}] 52\,$\micron$ line spectrum is $\lesssim$\,2, and that the fitted velocity width is so much larger than that of the [O\,{\sc i}] 63\,$\micron$ line (our most significant oxygen line detection), we additionally calculate the flux by summing over only the channels coincident with our [O\,{\sc i}] 63\,$\micron$ line (red dashed line and yellow shaded channels in Figure 1). It is this flux value, 6.8\,$\times$\,10$^{-18}$\,W\,m$^{-2}$, which we use in all calculations involving line ratios, since the flux excluded from this value originates within a different velocity component of 3C\,368.

With the exception of the [O\,{\sc iii}] 52\,$\micron$ line, the central line velocities, $\sim$ +\,200$-$300\,km\,s$^{-1}$ (calculated using $z$ = 1.131), and velocity widths,  $\sim$ 200$-$500\,km\,s$^{-1}$, calculated for the fine structure lines observed with \emph{Herschel/PACS} are very similar.

The flux values, along with central line velocities and FWHMs, for the higher signal to noise [Ne\,{\sc ii}] 12.81\,$\micron$ and [Ne\,{\sc iii}] 15.56\,$\micron$ lines observed with \emph{Spitzer/IRS} were obtained from the Infrared Database of Extragalactic Observables from Spitzer (IDEOS) \citep[][Spoon et al. in prep.]{Caballero2016}. The lower signal to noise [Ne\,{\sc v}] 14.32\,$\micron$ line flux was obtained by summing the line channels, exactly as done for the \emph{Herschel/PACS} spectra.

The central line velocities, determined by the Gaussian fits to the [Ne\,{\sc ii}] 12.81\,$\micron$ and [Ne\,{\sc iii}] 15.56\,$\micron$ lines, overlap with those calculated for the oxygen lines observed with \emph{Herschel/PACS} at the \textless\,2$\sigma$ uncertainty level, indicating that these neon lines likely originate within the same regions of 3C\,368 as do the oxygen lines.

\begin{figure*}
	\centering	
	\includegraphics[width=0.45\linewidth, height=5.4cm]{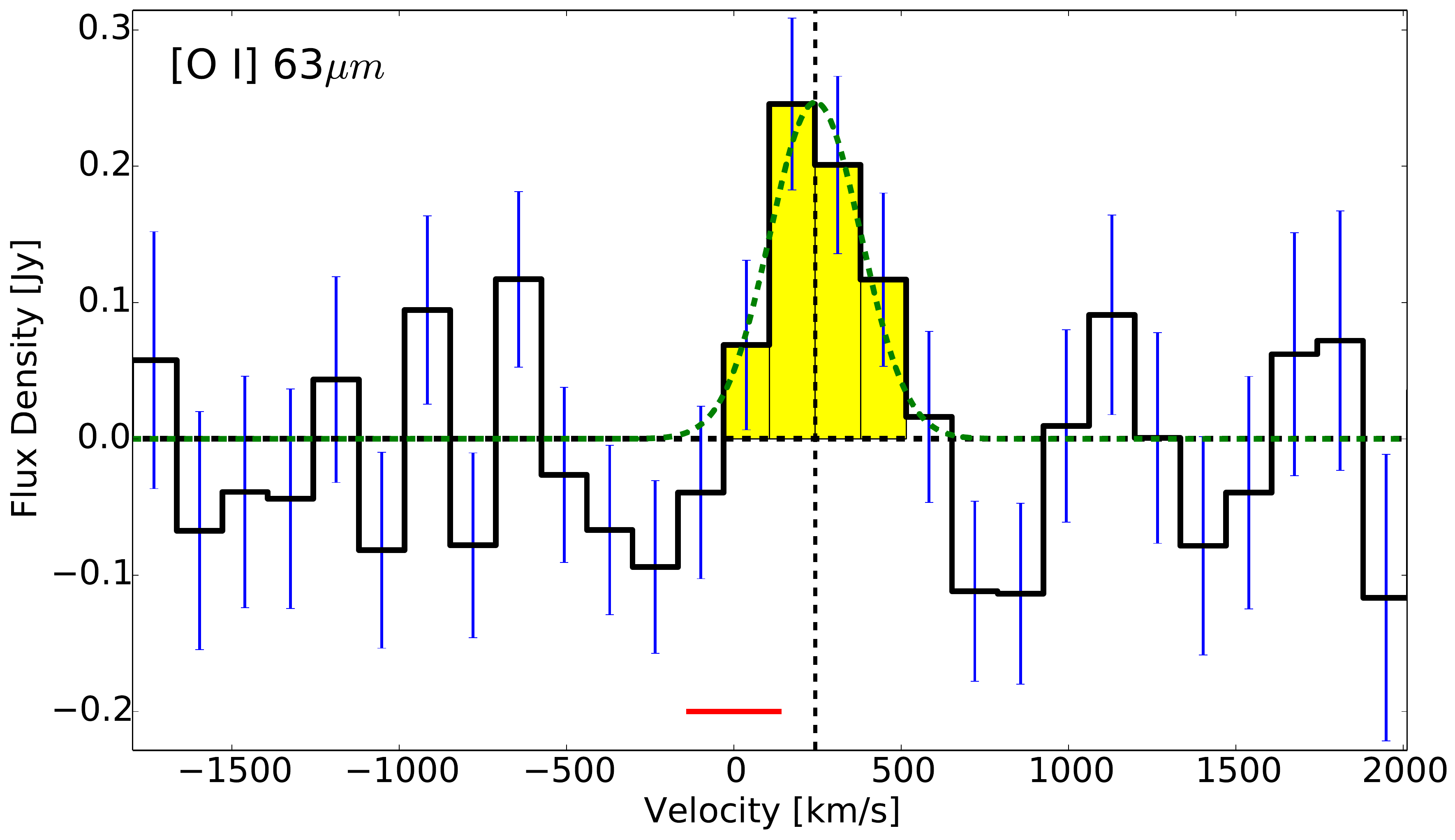}
	\centering
	\includegraphics[width=0.45\linewidth, height=5.4cm]{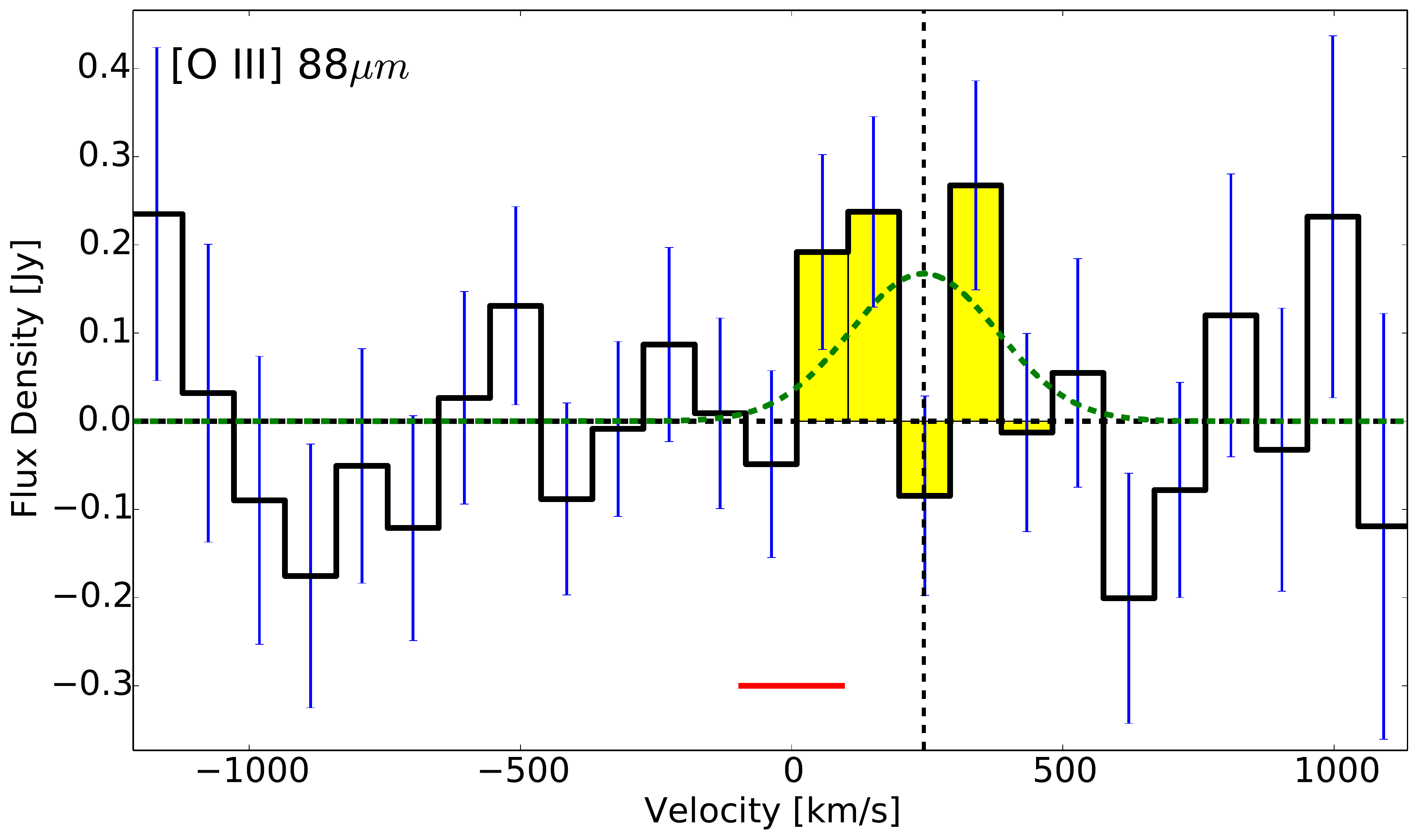}
	\centering
	\includegraphics[width=0.45\linewidth, height=5.4cm]{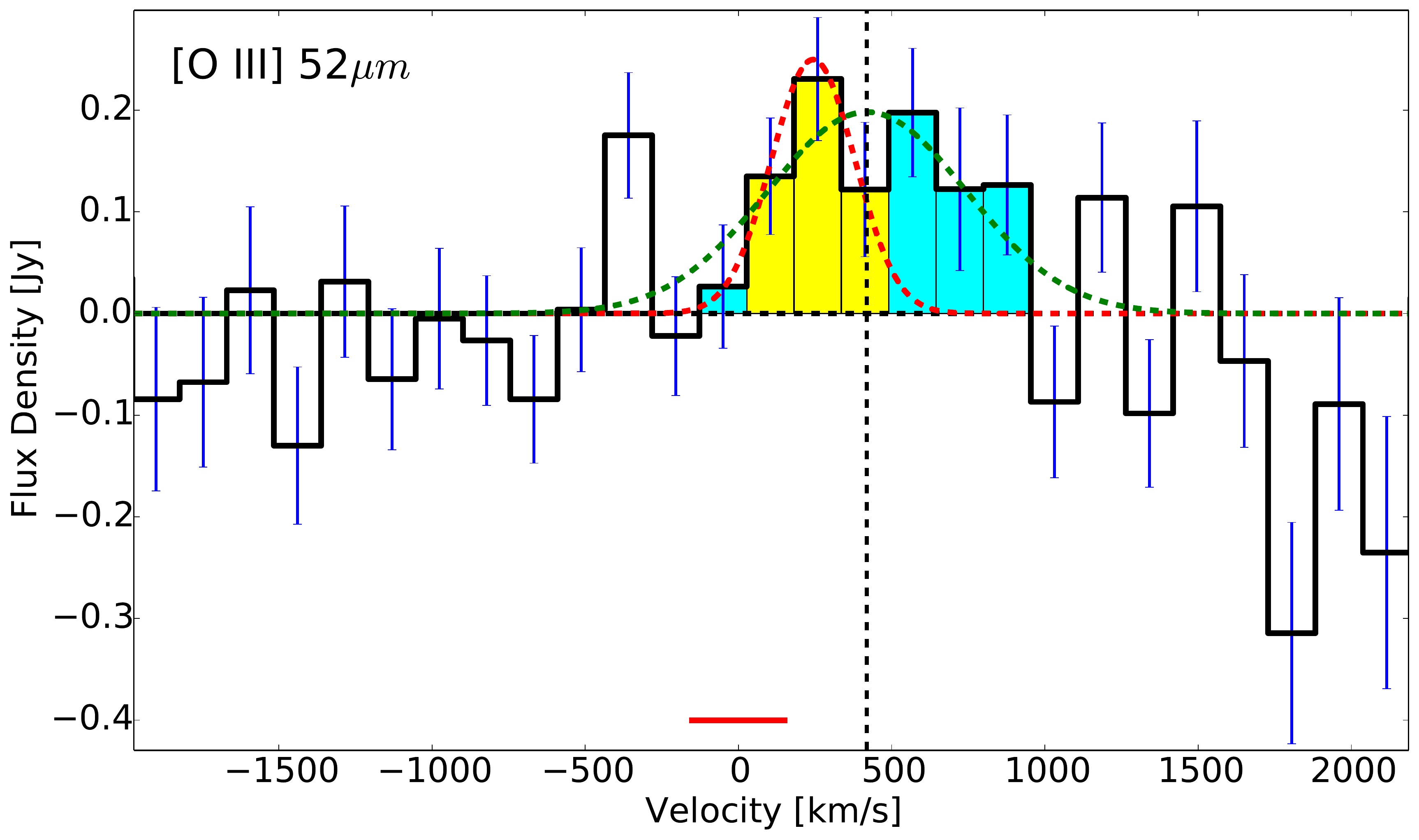}
	\centering
	\includegraphics[width=0.45\linewidth, height=5.4cm]{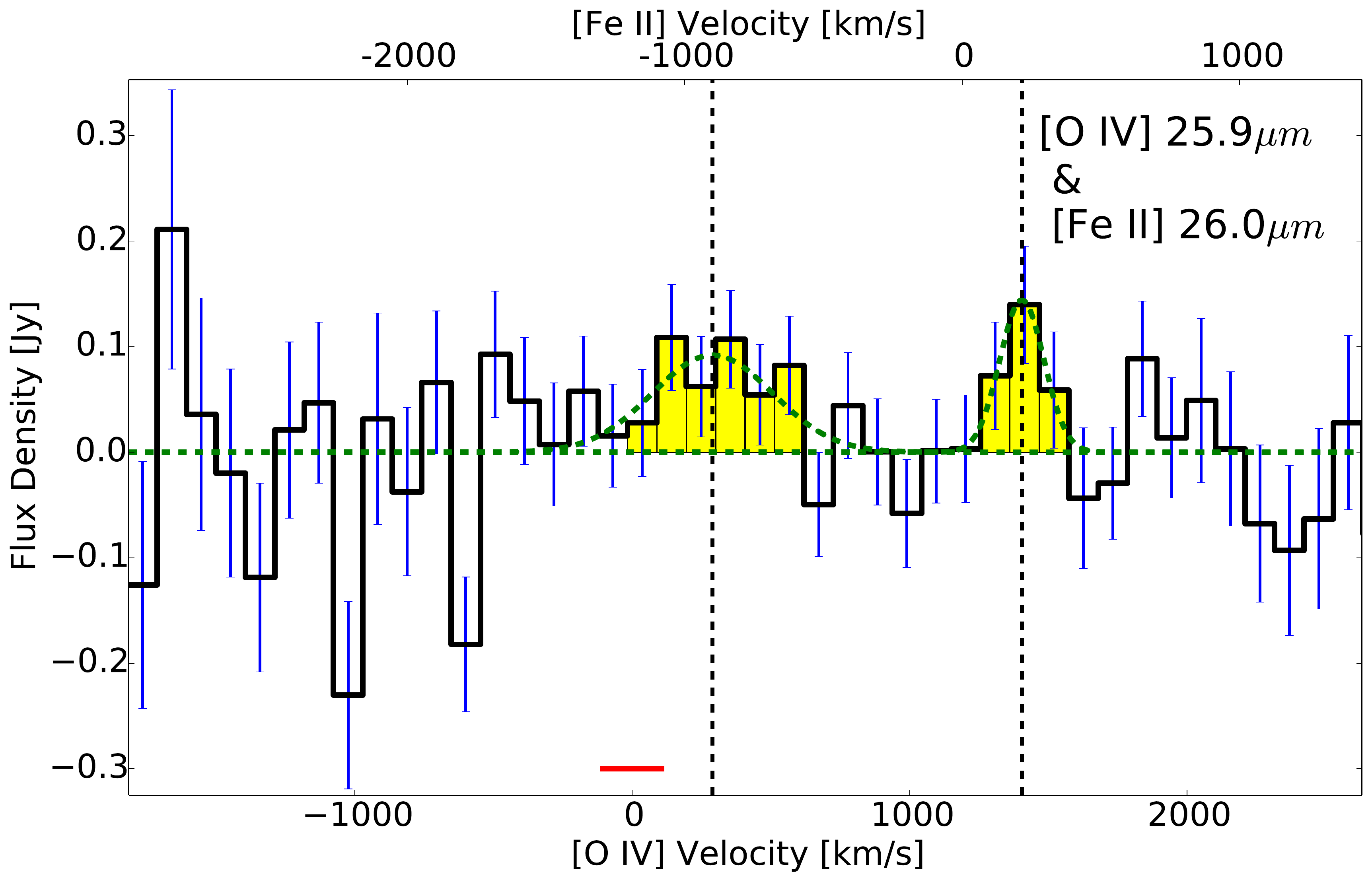}

\caption{Spectral lines observed in 3C\,368 with \emph{Herschel/PACS}: (Top Left) [O\,{\sc i}] 63\,$\micron$ line, spectral resolution $\sim$ 270\,km\,s$^{-1}$; (Top Right) [O\,{\sc iii}] 88\,$\micron$ line, spectral resolution $\sim$ 190\,km\,s$^{-1}$; (Bottom Left) [O\,{\sc iii}] 52\,$\micron$ line, spectral resolution $\sim$ 300\,km\,s$^{-1}$; (Bottom Right) [O\,{\sc iv}] 25.9\,$\micron$ and [Fe\,{\sc ii}] 26.0\,$\micron$ lines, spectral resolution $\sim$ 210\,km\,s$^{-1}$. Gaussian fits to the line profiles are shown by the green dashed lines, with the central velocities indicated by vertical black dashed lines ($v = 0$ corresponds to $z = 1.131$). The blue error bars represent the 1$\sigma$ statistical uncertainty in the flux measurement in each channel, and the red bar below each spectrum indicates the velocity resolution of that spectrum. The flux in the [O\,{\sc iii}] 52\,$\micron$ line is calculated in two different ways, one including both the yellow and cyan spectral channels, and the other using only the yellow spectral channels (see Section 3.1).}
\label{fig1}
\end{figure*}

\begin{figure}
\includegraphics[width=0.5\textwidth]{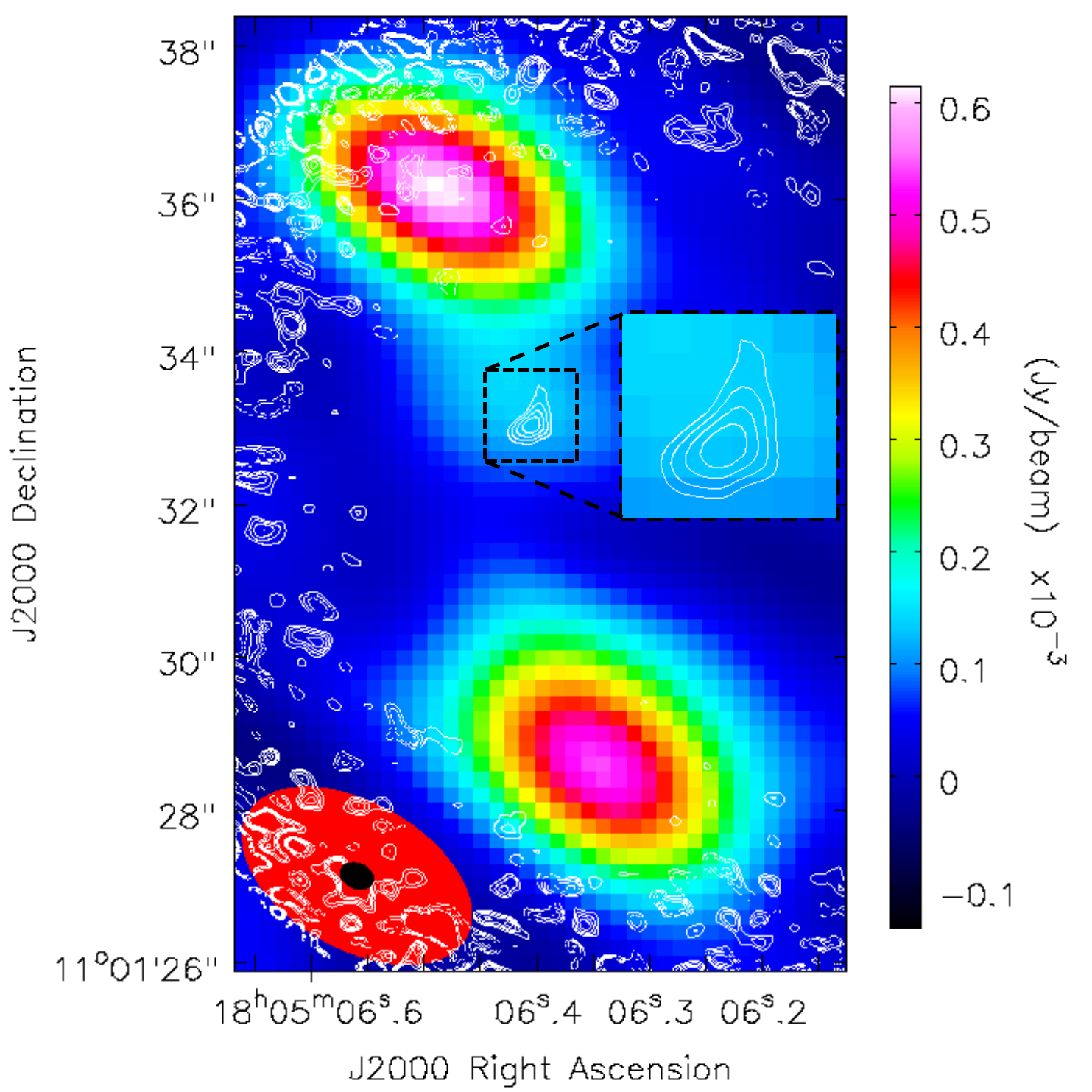}
\caption{ALMA observations of the 1.3\,mm (color map), and 205\,$\micron$ (contours), rest-frame continua in 3C\,368. The maps are primary-beam corrected. The contour levels are $\pm$3, $\pm$4, $\pm$5, and $\pm$6$\sigma$ (negative contours are dashed, 1$\sigma$ $=$ 0.4\,mJy\,beam$^{-1}$). The beams are shown at the bottom-left corner of the image, with the large (red) and small (black) ellipses for the 1.3\,mm and 205\,$\micron$ maps, respectively. The inset shows the 205\,$\micron$ rest-frame continuum from the core component of 3C\,368.}
\label{fig3}
\end{figure}

\begin{figure*}
	\includegraphics[width=\textwidth]{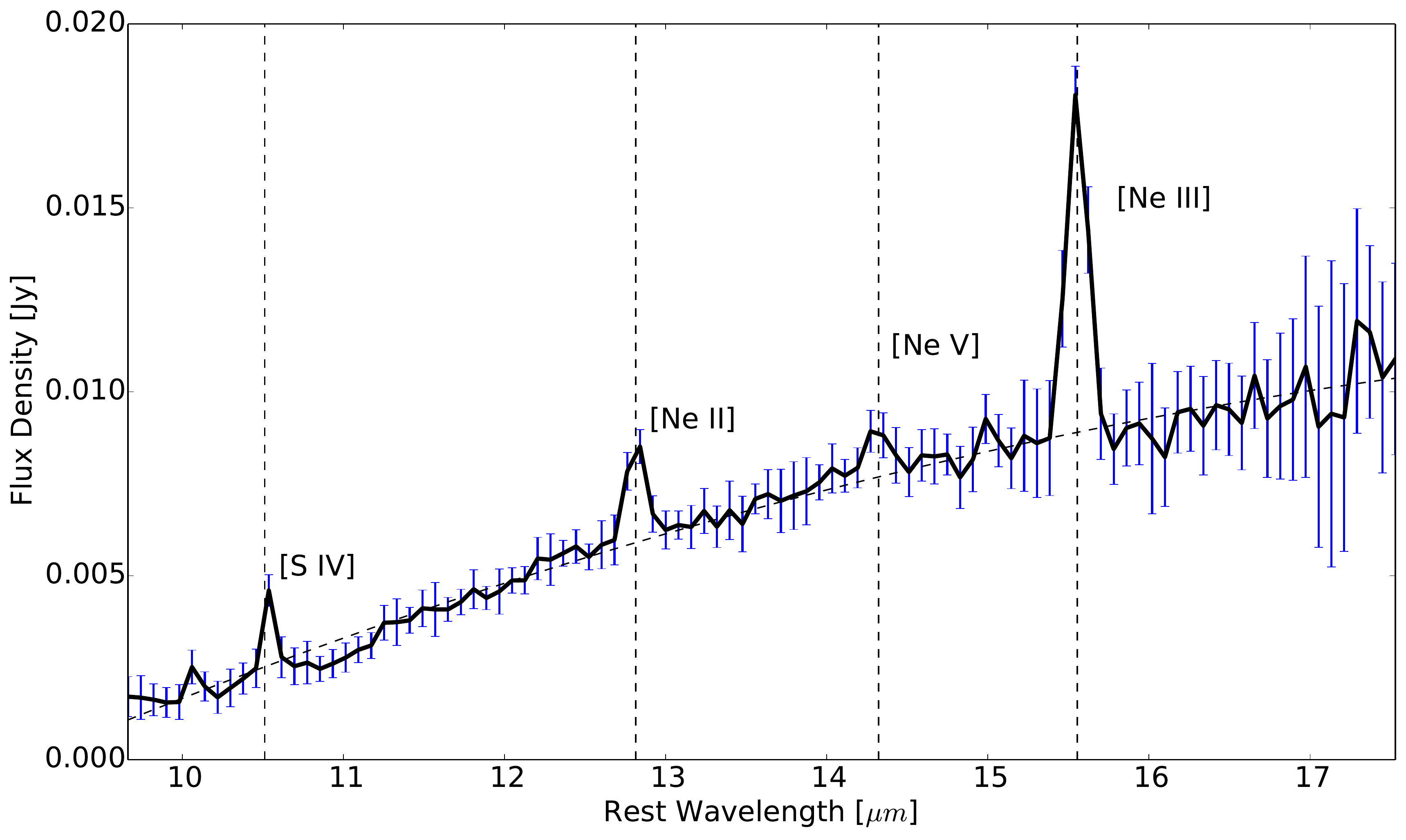}
	\caption{Spectral lines observed in 3C\,368 with \emph{Spitzer/IRS}. The positions of the various fine-structure lines are indicated by the vertical black dashed lines ($z = 1.131$). The blue error bars represent the 1$\sigma$ statistical uncertainty in the flux measurement in each channel. The spectral resolution ranges from $\sim60$--$120$ over the wavelength range of this observation.}
	\label{fig4}
\end{figure*}

\begin{figure}
\includegraphics[width=0.5\textwidth]{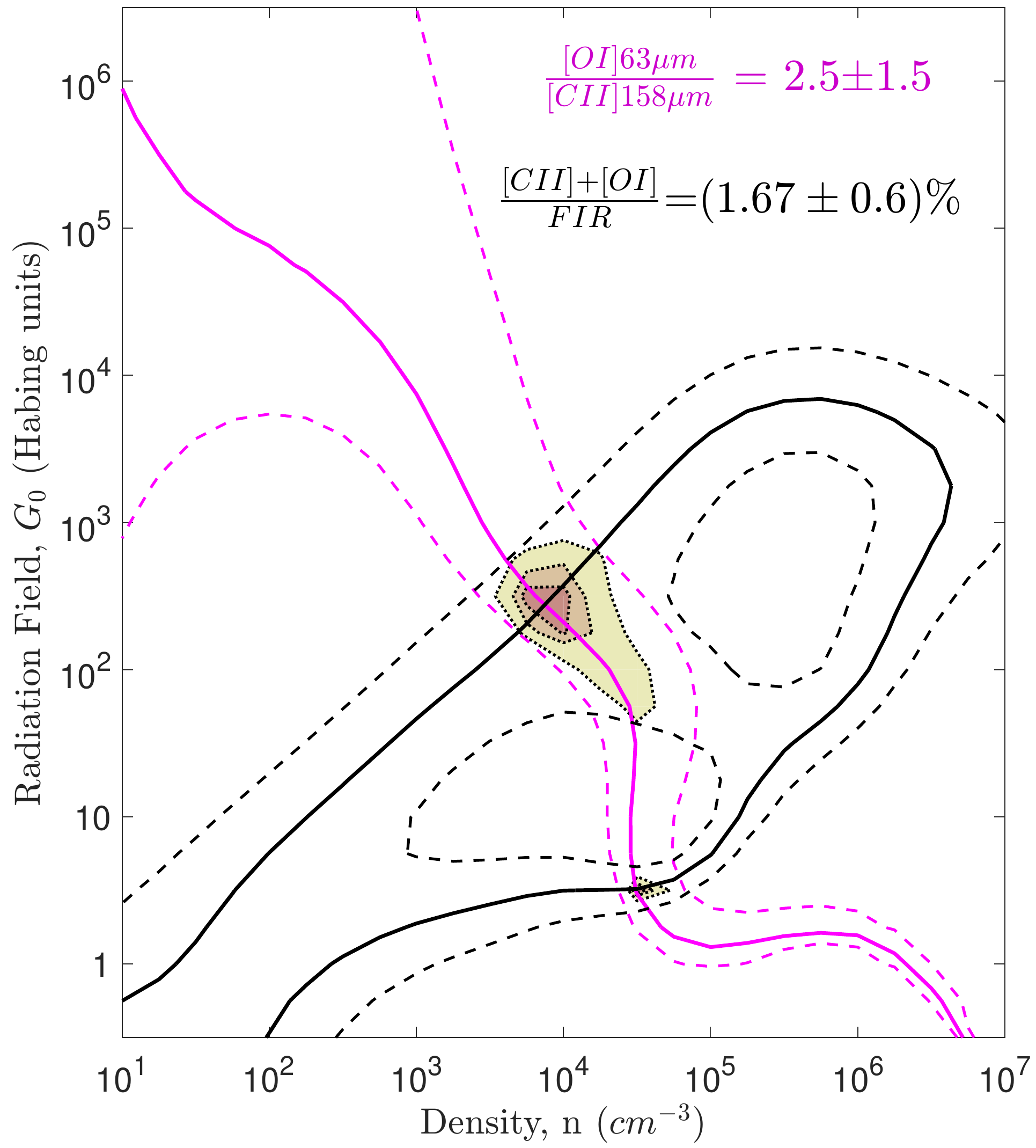}
\caption{A plot of the best-fit PDR parameters in 3C\,368, using the PDR Toolbox \citep{Kaufman2006}. The black (magenta) solid line traces the values of the radiation field, $G_0$, and density, n, allowed by the [C\,{\sc ii}]+[O\,{\sc i}]/FIR ([O\,{\sc i}]/[C\,{\sc ii}]) ratio, and is bounded by the uncertainty in the ratio (dashed lines). The three filled contours represent $\chi^2$ fitting values of 0.25, 0.5, and 1.0. The fit favors PDRs with $G_0$ $\sim$ 280 and n $\sim$ 7,500\,cm$^{-3}$. The low $G_0$ solution is disregarded because it would result in an emitting region with an unphysically large size.}
\label{fig5}
\end{figure}

\setlength{\tabcolsep}{0.2em}
\begin{deluxetable*}{lccccccccccc}
	\tabletypesize{\scriptsize}
	\centering
	\tablewidth{0pt}
	
	\tablehead{
		\colhead{Line} & \colhead{[Ne\,{\sc ii}]} & \colhead{[Ne\,{\sc v}]} & \colhead{[Ne\,{\sc iii}]} & \colhead{[O\,{\sc iv}]} & \colhead{[Fe\,{\sc ii}]} & \colhead{[O\,{\sc iii}]} & \colhead{[O\,{\sc i}]} & \colhead{[O\,{\sc iii}]} & \colhead{[C\,{\sc ii}]\tablenotemark{a}} & \colhead{[N\,{\sc ii}]} & \colhead{CO(2--1)} \\
		\colhead{ } & \colhead{12.81\,$\micron$} & \colhead{14.32\,$\micron$} & \colhead{15.56\,$\micron$} & \colhead{25.9\,$\micron$} & \colhead{26.0\,$\micron$} & \colhead{51.8\,$\micron$} & \colhead{63.2\,$\micron$} & \colhead{88.4\,$\micron$} & \colhead{158\,$\micron$} & \colhead{205\,$\micron$} & \colhead{1.3\,mm}
	}
	\startdata
Critical Density (cm$^{-3}$) & 7$\times$10$^5$\tablenotemark{b} & 3$\times$10$^4\tablenotemark{b}$ & 3$\times$10$^5$\tablenotemark{b} & 1$\times$10$^4$\tablenotemark{b} & 1$\times$10$^4$\tablenotemark{b} & 3.6$\times$10$^3$\tablenotemark{c} & 4.7$\times$10$^5$\tablenotemark{c}\tablenotemark{*} & 510\tablenotemark{c} & 2.8$\times$10$^3$\tablenotemark{c}\tablenotemark{*} & 48\tablenotemark{c} & 1.1$\times$10$^4$\tablenotemark{c}\tablenotemark{*}    \\
Flux ($10^{-18}$\,W\,m$^{-2}$) & 3.3\,$\pm$\,0.6 & 1.5 & 9.1\,$\pm$\,0.6 & 8.5 & 5.2 & 13.4\tablenotemark{d} & 6.4 & 3.0 & 5.1 & \textless\,0.0865 & \textless\,0.00073    \\
Significance ($\sigma$) & - & 2.9 & - & 3.8 & 2.9 & 5.5 & 5.0 & 2.4 & 6.4 & - & -    \\
Line Center (km s$^{-1}$) & -40\,$\pm$\,210 & - & -40\,$\pm$\,100 & 290\,$\pm$\,130 & 220\,$\pm$\,60 & 420\,$\pm$\,120 & 240\,$\pm$\,40 & 240\tablenotemark{e} & -140 & - & -    \\
Line Width (km s$^{-1}$) & 3100\,$\pm$\,700\tablenotemark{f} & - & 3000\,$\pm$\,200\tablenotemark{f} & 510\,$\pm$\,330 & 190\,$\pm$\,120 & 760\,$\pm$\,290 & 320\,$\pm$\,90 & 320\tablenotemark{e} & - & - & -    
\enddata
\tablecomments{Line fluxes were obtained by summing the flux over the line channels (shaded in Figure 1), while the line widths (FWHMs), positions ($v = 0$ corresponds to $z = 1.131$), and their associated uncertainties, were obtained from fitting Gaussian line profiles to the data. The higher signal-to-noise \emph{Spitzer/IRS} neon line fluxes were obtained from Gaussian fits to the line profiles. See Section 3.1.}

\tablenotetext{a}{\cite{Stacey2010CII}.}
\tablenotetext{b}{\cite{Cormier2012}.}
\tablenotetext{c}{\cite{Carilli2013}.}
\tablenotetext{d}{This value is the full [O\,{\sc iii}] 52\,$\micron$ line flux (sum of the yellow and cyan spectral bins in Figure 1). In all calculations involving line flux ratios, we sum only the spectral channels at the same velocity as the [O\,{\sc i}] 63\,$\micron$ line (yellow spectral bins in Figure 1). See Section 3.1 for further explanation.}
\tablenotetext{e}{The line position and FWHM of the [O\,{\sc iii}] 88\,$\micron$ line are not left as a free parameters in the Gaussian fitting. Instead, the prior from the [O\,{\sc i}] 63\,$\micron$ line is assumed. See Section 3.1 for further explanation.}
\tablenotetext{f}{The large FWHM of the fitted Ne lines is due to the resolution of the \emph{IRS} in the LL1 observing mode, and is not the physical width of these lines.}
\tablenotetext{*}{These critical densities correspond to lines originating from neutral gas regions.}

\end{deluxetable*}

\subsection{Neon Lines: Disentangling the AGN Contribution}

The mid-IR fine-structure lines in the \emph{Spitzer/IRS} spectral regime that arise from the three ionization states of neon are excellent probes of the hardness of the ambient UV radiation fields in galaxies. It takes 21.56\,eV photons to form Ne$^+$, 40.96\,eV photons to form Ne$^{++}$, and 97.11\,eV photons to form Ne$^{4+}$. Therefore, the 12.81\,$\micron$ [Ne\,{\sc ii}] line arises from H {\sc ii} regions formed by O/B stars, the 15.56\,$\micron$ [Ne\,{\sc iii}] line arises from H {\sc ii} regions formed by O stars, and the 14.32\,$\micron$ [Ne\,{\sc v}] line requires the very hard UV radiation fields found near AGN or planetary nebula created by hot white dwarfs. This series of lines is quite useful since extinction corrections between these lines are modest, and the lines have similar critical densities (Table 1). Furthermore, since neon is a noble gas, its gas phase abundance is not dependent on local conditions, since it is neither depleted onto grains nor incorporated into molecules.

[Ne\,{\sc v}] line emission is known to be associated with planetary nebulae exposed to the hard UV radiation of very young white dwarfs \citep[cf.][]{BernardSalas2001}, but on galactic scales, detectable [Ne\,{\sc v}] is predominantly associated with the NLR of AGN. However, detectable [Ne\,{\sc iii}] emission can arise from both the NLR of AGN and from H {\sc ii} regions formed by hot O stars. \cite{Gorjian2007} studied the neon fine-structure line emission from a variety of galaxies, including 77 3C radio sources, and found that the amount of [Ne\,{\sc iii}] line emission from the NLR of AGN is directly proportional to the [Ne\,{\sc v}] line emission. Therefore, a simple scaling law, based on the observed [Ne\,{\sc v}] line emission from a galaxy, can be applied to estimate the fraction of the observed [Ne\,{\sc iii}] line emission that arises from any central AGN.  We use this scaling law to``correct" the [Ne\,{\sc iii}] line emission for the AGN contribution before applying any H {\sc ii} region models to the star forming regions in 3C\,368.

Applying the relation of \cite{Gorjian2007},

\begin{equation}
\log \left( \frac{L_{\rm{Ne\sc{III}}}}{10^{33}\,W\,sr^{-1}} \right) = 0.30 + 0.89 \log \left( \frac{L_{\rm{Ne\sc{V}}}}{10^{33}\,W\,sr^{-1}} \right),
\end{equation}

\noindent to the observed luminosity of the [Ne\,{\sc v}] 14.32\,$\micron$ line, we estimate that 20\% of the observed [Ne\,{\sc iii}] emission originates from the AGN, leaving 80\% from star forming regions. This ``AGN-corrected" 80\% is the flux which we use in our star-formation-driven H {\sc ii} region models. We note that 3C\,368 is comparable to the higher luminosity sources used in the determination of this [Ne\,{\sc iii}]/[Ne\,{\sc v}] relation \citep[][Figure 1, right panel]{Gorjian2007}, where there are fewer sources and larger scatter in the trend, and so we also consider several limiting cases in the following analysis.

\subsection{H {\sc ii} Region Models}

Given the large number of fine-structure lines observed in 3C\,368, we can determine several properties of the H {\sc ii} regions within this source.

The ground-state-term level populations within the O$^{++}$ ion are density sensitive, so that the [O\,{\sc iii}] 52\,$\micron$/[O\,{\sc iii}] 88\,$\micron$ line ratio yields the ionized gas density. We find a line ratio of 2.3, which indicates H {\sc ii} region gas densities of n$_{e}$ $\sim$ 1000\,cm$^{-3}$ \citep[we here use the collision strengths from][]{Palay2012}.

Then, using the models of \cite{Rubin1985}, we can combine the [O\,{\sc iii}], [Ne\,{\sc ii}], and ``AGN-corrected" [Ne\,{\sc iii}] fine-structure lines to arrive at a consistent model for the H {\sc ii} regions in 3C\,368. We find that the fit favors regions of star formation heated by stars with effective temperature $\sim$ 37,000 K and with gas densities $\sim$ 1000\,cm$^{-3}$. The density is consistent with our estimates from the [O\,{\sc iii}] 52/88\,$\micron$ line ratio, and the ionization state is consistent with the neon line ratios, provided that neon is overabundant to oxygen in 3C\,368 by a factor of 3 compared with either the ``N" models, which have Orion Nebular abundances (O/H = 4.0\,$\times$\,10$^{-4}$ and N/H = 4.5\,$\times$\,10$^{-5}$), or the ``D" models, with all metals depleted by a factor of $\sqrt{10}$ from the N models.  

This factor of three overabundance is consistent with the recent results of \cite{Rubin2016}, who, using \emph{Spitzer} spectroscopy together with their ionization models, find higher neon to sulfur ratios ($\sim$12), for a wide variety of galaxies, than previously expected. This suggests that the cannonical neon to sulfur ratio ($\sim$3.7), which was used in the H {\sc ii} region models employed here, is too low by a factor of three or so. Since neon, sulfur, and oxygen are all primary elements, their abundances should scale together, and so our neon overabundance when compared to oxygen is consistent with this newly measured Ne/S ratio.

Given that the AGN contribution to the [Ne\,{\sc iii}] line is only $\sim$ 20\%, and considering the uncertainty in both the line fluxes and the scatter in the [Ne\,{\sc iii}]/[Ne\,{\sc v}] correlation (0.2 dex in log space), we consider both the limiting case of no AGN contribution to the [Ne\,{\sc iii}] line, and the case of the maximum allowed AGN contribution to the [Ne\,{\sc iii}] line ($\sim$ 40\%). In all cases, we find that the models favor regions of star formation heated by stars with effective temperature $\sim$ 37,000 K and with gas densities $\sim$ 1000\,cm$^{-3}$.

To retain the necessary effective stellar temperatures indicated by our modeling, at the observed far-IR luminosity in 3C\,368 \citep[2.0\,$\times$\,$10^{12}$ $L_{\odot}$,][]{Podigachoski2015}, we require a starburst with the equivalent of $\sim$ 1.2\,$\times$\,$10^{7}$ O8 V stars \citep{Vacca1996}. Such a starburst would have an age of $\sim$ 6.5 Myr \citep{Meynet2003}, as constrained by the stellar lifetimes of the most massive stars. Using the 73\,kpc radio source size of 3C\,368 from \cite{Best1998}, and a range of jet propagation speeds from 0.03 $-$ 0.3\,c \citep{King2015}, we estimate the duration of the latest episode of AGN activity to be between 0.4 $-$ 4\,Myr, comparable in time to the age of the starburst.

This concurrence lends further support to the possibility that we may be witnessing AGN-driven star formation in 3C\,368, as has been suggested previously from the alignment of the radio and optical axes, which trace AGN jets and stars, respectively, seen in many of the 3CR galaxies, including 3C\,368 \citep{McCarthy1987}. In addition to the optical continuum, \cite{Chambers1988} find the 2.2 $\micron$ continuum, also associated with stellar emission, to be aligned with both of the other axes.

It has also been previously suggested that 3C\,368 may be undergoing a major merger \citep[][]{Djorgovski1987}, providing another possible mechanism for triggering both black hole accretion and star formation simultaneously. While we cannot rule out this possibility, the [C\,{\sc ii}]/$F_{\rm{FIR}}$ ratio of 0.5\% in 3C\,368 (see Sections 3.5 \& 3.6) is more than an order of magnitude larger than the typical values found in the local Ultraluminous Infrared Galaxies (ULIRGs), which are predominantly powered by major-merger-driven starbursts \citep[e.g.,][]{Luhman1998}.

The physical parameters derived from our H {\sc ii} region modeling also explain our non-detection of the [N\,{\sc ii}] 205\,$\micron$ line. At such large effective temperatures, \cite{Rubin1985} calculate that $\sim$ 86$\%$ of the nitrogen contained within the H {\sc ii} regions of 3C\,368 would be in the N$^{++}$ state. Unfortunately, the [N\,{\sc iii}] 57\,$\micron$ line is redshifted into an absorption band in the atmosphere for 3C\,368 with $z$ = 1.131 ($\sim$ 122\,$\micron$), making observations impossible from the ground, and even beyond the reach of airborne facilities. These observations will have to wait for the next generation of far-IR space telescopes, such as the Space Infrared Telescope for Cosmology and Astrophysics (SPICA) and the Origins Space Telescope (OST).

\subsection{Bounds on the N/O Ratio}

Following \cite{Ferkinhoff2010}, we can estimate the minimum mass of oxygen in the O$^{++}$ state in 3C\,368, from the luminosity in the oxygen fine-structure lines:

\begin{equation}
M_{\rm{min}}(O^{++}) = \frac{F_{ul} 4 \pi D_{L}^{2} m_O}{\frac{g_l}{g_t} A_{ul} h \nu_{ul}} \rm{,}
\end{equation}

\noindent where $F_{ul}$ is the flux in the fine-structure line between the upper ($u$) state and the lower ($l$) state, $D_L$ is the luminosity distance (7.735\,Gpc), $m_O$ is the mass of an oxygen atom, $g_u$ and $g_l$ are the statistical weights of the upper and lower states, respectively, $g_t$ is the partition function (the sum of the statistical weights of all relevant states available to the O$^{++}$ ion at T $=$ 8,000 K), $A_{ul}$ is the Einstein coefficient for the relevant transition \citep[2.6\,$\times$\,10$^{-5}$\,s$^{-1}$ for the {[}O\,{\sc iii}{]} 88\,$\micron$ line,][]{Carilli2013}, and $\nu_{ul}$ the frequency of that transition. Using the [O\,{\sc iii}] 88\,$\micron$ line, we obtain a minimum $O^{++}$ mass of 4.4\,$\times$\,10$^6$  $M_{\odot}$.

Similarly, we can use our 3$\sigma$ upper limit for the [N\,{\sc ii}] 205\,$\micron$ line flux \citep[$A_{ul}$ $=$ 2.1\,$\times$\,10$^{-6}$\,s$^{-1}$,][]{Carilli2013} to put a bound on the mass of N$^+$ in 3C\,368:

\begin{equation}
M_{\rm{min}}(N^+) < \frac{F_{ul} 4 \pi D_{L}^{2} m_N}{\frac{g_l}{g_t} A_{ul} h \nu_{ul}} \rm{.}
\end{equation}

\noindent We obtain a limit of \textless 3.2\,$\times$\,10$^6$ $M_{\odot}$.

We use the models of \cite{Rubin1985} to calculate the total mass in oxygen and nitrogen by scaling from the fraction in the ionization state that we observe back to the total abundance of the element in question. 

With $\sim$ 84$\%$ of the oxygen in the O$^{++}$ state, and $\sim$ 14$\%$ of the nitrogen in the N$^{+}$ state, within the H {\sc ii} regions of 3C\,368, we calculate masses of 5.3\,$\times$\,10$^6$  $M_{\odot}$ and \textless\, 2.3\,$\times$\,10$^7$  $M_{\odot}$, for oxygen and nitrogen respectively.

Using these values, we find an upper limit for the N/O ratio in 3C\,368 of 5.0. \cite{Asplund2009} report a solar N/O abundance ratio of 0.126. While the result which we obtain is consistent with the solar N/O ratio, it also allows for enhanced nitrogen relative to oxygen. 

We are continuing our campaign to detect the [N\,{\sc ii}] 205\,$\micron$ line in 3C\,368, and were recently granted ALMA time to push our sensitivity deeper in this source. These new observations will effectively double our on-source integration time, improving our sensitivity.

\subsection{PDRs}

In addition to the lines associated with H {\sc ii} regions, we have also detected the [O\,{\sc i}] 63\,$\micron$ line associated with photodissociation regions (PDRs). [O\,{\sc i}] and [C\,{\sc ii}] are the two dominant coolants of PDRs, so that the sum of the line fluxes divided by the far-IR flux (which tracks the impinging FUV (6 $-$ 13.6\,eV) flux) tracks the efficiency of photoelectric heating.  Parameterizing the FUV in terms of the local (Habing) interstellar radiation field, the photoelectric heating is sensitive to the ratio of FUV field strength to gas density, G$_{0}$/n.  With different critical densities the line ratio constrains the PDR gas density, so that the line intensities and their ratio constrain G$_{0}$ and n independently.

We use our prior detection of [C\,{\sc ii}] \citep{Stacey2010CII}, together with  the PDR Toolbox \citep{Kaufman2006}, for our analysis. This software fits input line flux ratios to a bank of line ratios calculated using a radiative transfer code that assumes a plane
parallel geometry for the gas.

Before using this software, we correct for the different optical depths of the lines used in modeling the PDRs. The [O\,{\sc i}] 63\,$\micron$ line is likely to be optically thick for PDRs with A$_V$ of the order of a few \citep[e.g.,][]{Stacey1983, Tielens1985}. We, therefore, multiply the [O\,{\sc i}] 63\,$\micron$ line flux by a factor of two, in order to account for its opacity \citep[as in][]{Stacey2010PDR}, taking the uncertainty in this correction to be equal to the correction which we applied.

Due to the lesser abundance of C$^+$, and the greater populations in its excited levels, the [C\,{\sc ii}] line is expected to have significantly smaller optical depth than the [O\,{\sc i}] line. Observations of the hyperfine structure lines of $^{13}$C$^+$ indicate optical depths less than or of the order of 0.5 to 1.5 for galactic PDRs \citep[cf.][]{Stacey1991, Boreiko1988, Ossenkopf2013}. PDR models indicate optical depths of the same order \citep{Kaufman1999}. Therefore, we make no correction to the [C\,{\sc ii}] 158\,$\micron$ line for opacity effects.

We perform an additional correction to remove any contributions to the [C\,{\sc ii}] 158\,$\micron$ line emission originating from H {\sc ii} regions, before performing our PDR analysis. With an ionizing energy of only 11.26\,eV, ionized carbon can be found both in the neutral hydrogen phases of the ISM (13.6\,eV ionizing energy), and also in the ionized gas phase where species like ionized nitrogen (14.5\,eV) exclusively reside \citep{Oberst2006}. Since the [N\,{\sc ii}] 205\,$\micron$ and [C\,{\sc ii}] 158\,$\micron$ lines have very similar critical densities in ionized gas regions, 48\,cm$^{-3}$ and 50\,cm$^{-3}$, respectively, their ratio can be used to calculate the percentage of C$^+$ coming from H {\sc ii} regions \citep{Oberst2006}. We calculate a flux ratio of [C\,{\sc ii}] 158\,$\micron$ / [N\,{\sc ii}] 205\,$\micron$  \textgreater\,60. \cite{Oberst2006} find that if all of the C$^+$ observed were to come from H {\sc ii} regions, this ratio would vary from 3.1 for a low density gas to 4.3 for a high density gas, so that \textless\,7\% of the [C\,{\sc ii}] 158\,$\micron$ line flux in 3C\,368 comes from within H {\sc ii} regions. Since this contribution is very small and within the uncertainties, we make no correction to the [C\,{\sc ii}] 158\,$\micron$ line flux for our PDR analysis.

Finally, we correct the observed FIR luminosity to include only the component produced by the UV flux from young stars which has been reprocessed by the dust into the FIR. We adopt the fit from \cite{Podigachoski2015}, who model the SED of 3C\,368 with three components: one for the AGN-fueled warm dust component \citep[using a library of torus models from][]{Honig2010}, one black-body for the optical/NIR stellar emission, and one gray-body in the FIR/sub-mm with the dust emissivity index ($\beta$) as a free parameter. We use the luminosity in this last component for our PDR analysis. \cite{Podigachoski2015} also use this FIR luminosity to calculate a star formation rate in 3C\,368, obtaining a value of 350\,$M_\odot$\,yr$^{-1}$.

Inserting these three observations (see Table 2), the corrected [C\,{\sc ii}] 158\,$\micron$ and [O\,{\sc i}] 63\,$\micron$ line intensities, and the fitted FIR luminosity, into the PDR Toolbox software \citep{Kaufman2006}, we obtain best fits for PDRs with $G_0$ $\sim$ 280 and n $\sim$ 7,500\,cm$^{-3}$ (see Figure 4). Using the scaling relations found in \cite{Wolfire1990}, and adopting an FIR luminosity of 2.0\,$\times$\,$10^{12}$ $L_{\odot}$ \citep{Podigachoski2015}, we calculate that these PDRs are extended over 1.9 $-$ 4.8\,kpc, consistent with our CASA 2-D Gaussian fit to the source from the rest-frame 205$\micron$ continuum map ($\sim$1.0\,$\times$\,5.9\,kpc, see Figure 2).

One potential caveat for our PDR modeling results is that the observed [C\,{\sc ii}] and [O\,{\sc i}] lines may originate within the X-ray dominated region (XDR) surrounding the AGN of 3C\,368 \citep[e.\/g.\/,][]{Meijerink2007}, with the [O\,{\sc iii}] lines coming from the closer-in NLR \citep[e.\/g.\/,][]{Ferkinhoff2010}. Given that we can form a consistent picture of star formation in 3C\,368 using our observations, and that optical fine-structure lines have also been observed to be extended over several arc seconds around the central source \citep[e.\/g.\/,][]{Hammer1991, Meisenheimer1992, Jackson1997, Best2000}, the theory of large-scale star formation in 3C\,368 seems more plausible.
 
This theory can be tested by spatially resolving the [C\,{\sc ii}] line emission in 3C\,368. Extended [C\,{\sc ii}] emission may confirm the existence of large-scale star forming regions, while confined emission may suggest an AGN origin. We have been awarded ALMA time to carry out these very observations (see Future Observations section).

\begin{deluxetable}{lccccccc}
\tabletypesize{\scriptsize}
\tablecaption{PDR Modeling Line Corrections}
\tablewidth{0pt}
\tablehead{
\colhead{Line} & \colhead{[O\,{\sc i}]} & \colhead{[C\,{\sc ii}]\tablenotemark{a}} &
\colhead{$F_{\rm{FIR}}$}\tablenotemark{b} & \\
\colhead{ } & \colhead{63.2\,$\micron$} & \colhead{158\,$\micron$} & \colhead{ } & \vspace{-0.1cm}
}
\startdata
Uncorrected Flux ($10^{-18}$\,W\,m$^{-2}$) & 6.4 & 5.1 & 1075.0    \\
Corrected Flux ($10^{-18}$\,W\,m$^{-2}$) & 12.8 & 5.1 & 1075.0    \\
Uncertainty ($10^{-18}$\,W\,m$^{-2}$) & 6.4\tablenotemark{c} & 1.5 & 107.5
\enddata

\tablenotetext{a}{\cite{Stacey2010CII}.}
\tablenotetext{b}{\cite{Podigachoski2015}.}
\tablenotetext{c}{We take the uncertainty in the [O\,{\sc i}] 63\,$\micron$ line to be equal to the opacity correction which we applied (see Section 3.5).}

\end{deluxetable}

\subsection{Where's the CO?}

While the fine-structure lines from PDRs allow us to model these regions in 3C\,368, the complete picture of star formation can only come from including molecular gas observations as well. We have attempted to observe the CO(2--1) transition with both ALMA and PdBI, and like others in the past \citep[e.\/g.\/,][]{Evans1996}, we have not detected any emission. Our ALMA observations probe down to a sensitivity of 134\,$\mu$Jy\,beam$^{-1}$, over an assumed line width of 500\,km\,s$^{-1}$.

Assuming a ratio of [C\,{\sc ii}] to CO(1--0) of $\sim$ 4,100, which is typical of normal metallicity star forming regions in the Milky Way and starburst galaxies \citep{Stacey1991}, and a CO(2--1) to CO(1--0) flux ratio of 7.2 (90$\%$ of the high-temperature thermalized value), we would expect a [C\,{\sc ii}] to CO(2--1) flux ratio of $\sim$ 570 in 3C\,368. The 3$\sigma$ CO upper limit from our ALMA observations, however, yields a limit of [C\,{\sc ii}]/CO(2--1) \textgreater\,7,000, which is more than 12\,$\times$ the expected value. 

Following \cite{Scoville2016}, we can estimate the total molecular gas mass in 3C\,368 using the 850\,$\micron$ SCUBA observations from \cite{Archibald2001}. Assuming a dust temperature of 35\,K, consistent with the SED fit from \cite{Podigachoski2015}, we obtain a value of 1.57\,$\times$\,10$^{11}$ M$_{\odot}$. Then, using our upper limit for the CO(2--1) line, and assuming that the gas is fully thermalized, we obtain a 3\,$\sigma$ limit for L$^{'}_{CO}$\,\textless \,3.45\,$\times$\,10$^{9}$\,K\,km\,s$^{-1}$\,pc$^{2}$. Using these values, we estimate an $\alpha_{CO}$\, \textgreater \,45.5\,M$_{\odot}$\,(K\,km\,s$^{-1}$\,pc$^{2}$)$^{-1}$ in 3C\,368. This $\alpha_{CO}$ value is an order of magnitude larger than the typical range of values quoted in the literature, $\sim$0.8$-$4.3 \citep[e.g.,][]{Solomon2005, Bolatto2013}.

Previous attempts to detect molecular gas using the CO(1--0) line in high redshift radio galaxies have been largely unsuccessful \citep[e.g.,][]{Emonts2014}. Our limit of L$^{'}$$_{CO}$\,\textless\,3.45\,$\times$\,10$^{9}$\,K\,km\,s$^{-1}$\,pc$^{2}$ is a factor of $\sim$6.4 deeper than the lowest limit reported in \cite{Emonts2014} (L$^{'}$$_{CO}$\,\textless\,2.2\,$\times$\,10$^{10}$\,K\,km\,s$^{-1}$\,pc$^{2}$).

Comparing the observed [C\,{\sc ii}]/$F_{\rm{FIR}}$ and upper-limit CO(1--0)/$F_{\rm{FIR}}$ ratios for 3C\,368 to those of other sources \citep[][Figure 3]{Hailey-Dunsheath2010}, we find that 3C\,368 has a higher [C\,{\sc ii}]/$F_{\rm{FIR}}$ ratio ($\sim$ 0.5\%) than either the starburst nuclei or galactic star forming regions, and deficient CO for a source with this [C\,{\sc ii}] flux. There are several factors which could explain this anomalously high [C\,{\sc ii}]/CO(2--1) line ratio, including the metallicity, age, and degree of fractionation of the molecular clouds in 3C\,368.

Moving toward the core of a molecular cloud from its surface, a transition is made from C$^+$ to CO as carbon-ionizing and CO-dissociating photons are extincted by dust. This transition occurs at A$_V$ $\sim$ 3 \citep{Hollenbach1999}. If the dust-to-gas ratio follows metallicity, then it follows that the penetration depth of these carbon-ionizing photons can be very large in a low-metallicity ISM, making the CO core relatively small compared to the C$^+$ envelope, and leading to large [C\,{\sc ii}]/CO ratios \citep[e.\/g.\/,][]{Maloney1988, Stacey1991}. While the CO emitting core of a molecular cloud retreats as the metallicity decreases, the size of the molecular region, defined by the transition from H to H$_2$, does not. This transition from H to H$_2$ is set by the self shielding of H$_2$ molecules, and is therefore metallicity independent. CO would therefore be a poor tracer of the total molecular gas in low-metallicity molecular clouds, which could still be present in the near absence of CO \citep[e.\/g.\/,][]{Stacey1991, Poglitsch1995, Madden1997}.

However, the low metallicity case is unlikely for 3C\,368, which has a stellar mass of $\sim$ 3.6\,$\times$\,10$^{11}$\,$M_{\odot}$ \citep{Best1998stellarmass}. At the current star formation rate, it would take $\sim$ 10$^9$\,years to accumulate such a stellar population. Assuming that each star formation episode lasts $\sim$ 10$^7$ $-$ 10$^8$\,years, 3C\,368 would have been through $\sim$ 10 $-$ 100\,cycles of star formation, making a low metallicity ISM seem improbable. Additionally, 3C\,368 lies on the ``galaxy main sequence" for a source at redshift 1.131 \citep[using the definition of][]{Genzel2015}, furthering the idea that a low metallicity ISM is not responsible for the lack of CO in this case.

A different scenario which could be used to explain the lack of CO is that the ISM in 3C\,368 may be highly fractionated into small cloudlets. Such a fractioned ISM would have modest extinction to the core of individual cloudlets, allowing for considerable C$^+$ emission with relatively little CO emission.

Since the upper limit from our CO observations show that the line is suppressed from the expected value by more than 12\,$\times$, the CO emitting cores of the molecular clouds in 3C\,368 must correspondingly be more than $\sqrt{12}$\,$\times$ smaller in radius than the entire clouds. Taking A$_V$ $\sim$ 3, or a corresponding column density of 6\,$\times$\,10$^{21}$\,cm$^{-2}$, as the CO depth, and our fitted PDR density of $\sim$ 7,500\,cm$^{-3}$, we calculate that the clouds in 3C\,368 must have size $\sim$\,0.3\,pc, smaller than the sizes derived for clouds in the highly disturbed ISM of M82 by \cite{Lord1996} (0.4 $-$ 1.0\,pc), while the CO emitting cores of these clouds must have size \textless\,0.1\,pc.

Yet another possible explanation for the lack of CO in 3C\,368 is that the molecular clouds in the galaxy may be chemically young. \cite{Glover2012} have shown that C$^+$ recombines with free electrons to form C$^0$ in less than $\sim$ 0.1\,Myrs, while the subsequent formation of CO from the neutral atoms can take significantly longer, $\sim$ 1$-$3\,Myrs. It could be that we are observing these molecular clouds before they have had the chance to form an appreciable amount of CO. 

The work of \cite{Storzer1997} suggests that a chemically young ISM may be possible here. Cloudlets within a clumpy ISM, moving relative to one other, can periodically shield each other from the intense UV radiation field present in the ISM. In this scenario, the [C\,{\sc ii}] which we observe would originate from the surfaces of dense cloudlets exposed to the intense UV field. The CO emitting cores of the cloudlets would retreat toward the center each time they were exposed to the full UV field, and then take an appreciable time to reform after the next shielding event. This effect would be particularly pronounced if the cloudlets were small, and hence much of their volume was penetrated by UV photons between shielding events, as seems to be the case here. A testable prediction of this model would be the existence of considerable [C\,{\sc i}] emission, since C$^+$ recombines quickly after the shielding event is initiated. For this reason, we plan to do follow up observations of 3C\,368 in [C\,{\sc i}], in order to try and resolve the mystery of the missing CO.

\section{Future Observations}

In an effort to answer some of the remaining questions pertaining to 3C\,368, we have been awarded ALMA time to observe the [C\,{\sc ii}] 158\,$\micron$ line, at 0$\farcs$2 spatial resolution, to determine the extent of star formation in this source. The models which we have employed here predict star formation extended over kilo-parsec scales, which, if correct, will appear in the ALMA observations as an extended source over $\sim$10\,beams.

In addition to the ionized carbon line, we have also been awarded time to observe the [C\,{\sc i}] 609\,$\micron$ line in 3C\,368 (the 370\,$\micron$ line is blocked by the atmosphere). We expect that if the molecular clouds in the galaxy are indeed chemically young, the neutral carbon line will be very bright when compared to our CO line upper limit, since the CO would not have had sufficient time to form.

\section{Conclusions}

We have observed the fine-structure lines constituting the ``Oxygen Toolkit" in 3C\,368, including the [O\,{\sc i}] 63\,$\micron$, [O\,{\sc iii}] 52\,$\micron$ and 88\,$\micron$, and [O\,{\sc iv}] 26\,$\micron$ lines. Using these oxygen lines together with archival fine-structure neon lines in the mid-infrared, we have modeled the H {\sc ii} regions in 3C\,368, using the models of \cite{Rubin1985}, and found them to be ionized by starbursts headed by O8 stars, with an age of $\sim$ 6.5\,Myr. This age is comparable to the estimated age of the latest episode of AGN activity in 3C\,368, determined by the propagation of the radio lobes.

Given the rate of star formation that we are witnessing in 3C\,368, with the alignment of the optical and radio axes, and estimated age of both the starburst and most recent epoch of AGN activity, this source is a strong candidate for further study in the field of galaxy-BH coevolution.

We have also modeled the PDRs in 3C\,368, and found them to be consistent with star formation spread over kilo-parsec distance scales. However, we have not detected any CO in 3C\,368. Our 3$\sigma$ limit for the CO(2--1) line is \textless 0.201\,Jy\,km\,s$^{-1}$, or equivalently \textless 7.3\,$\times$\,10$^{-22}$\,W\,m$^{-2}$, a level 12\,$\times$ smaller than the value expected from standard PDR models. The lack of CO emission may be due to a low-metallicity, highly fractionated, or possibly chemically young ISM. We find that the most likely explanation is that the ISM is highly fractionated into small clouds, perhaps as a result of the interactions between AGN winds and the star forming molecular clouds, making it appear chemically young. Future ALMA observations will enable us to test this theory of large-scale star formation in the highly fractionated ISM of 3C\,368.

\section*{Acknowledgments}

We thank the anonymous referee for the insightful comments and suggestions which helped to improve this manuscript. We additionally thank T. K. Daisy Leung and A. Gowardhan for help with preparing this manuscript in LaTeX. ZEUS observations were supported by NSF grant AST-01109476. C.L. acknowledges support from an NRAO Student Observing Support Award, SOSPA3-011, and a New York Space Grant Award. D.B. acknowledges partial support from ALMA-CONICYT FUND No. 31140010. D.R. acknowledges support from the National Science Foundation under grant number AST-1614213 to Cornell University.

This paper makes use of the following ALMA data: ADS/JAO.ALMA$\#$2012.1.00426.S and ADS/JAO.ALMA$\#$2013.1.01223.S. ALMA is a partnership of ESO (representing its member states), NSF (USA) and NINS (Japan), together with NRC (Canada) and NSC and ASIAA (Taiwan) and KASI (Republic of Korea), in cooperation with the Republic of Chile. The Joint ALMA Observatory is operated by ESO, AUI/NRAO and NAOJ.

This work is based in part on observations made with the \emph{Spitzer Space Telescope}, which is operated by the Jet Propulsion Laboratory, California Institute of Technology under a contract with NASA.

\end{document}